\documentclass[prl,showpacs,preprintnumbers,amsmath,amssymb]{revtex4}
\usepackage[dvips]{graphicx}
\usepackage{dcolumn}

\begin{document}


\noindent{\bf 
Comment on "Spatial line nodes and fractional vortex pairs in the Fulde-Ferrell-Larkin-Ovchinnikov phase":}





\bigskip

\ Recent observations \cite{Bianchi,Watanabe} of a high field superconducting (SC) phase in CeCoIn$_5$ have led to intense interest \cite{AI,I2} in modulated {\it vortex} phases induced by the paramagnetic depairing. Authors of Ref.\cite{Agter} have discussed a possibility that, as a candidate of the high field state in CeCoIn$_5$, a crisscrossing lattice of {\it integer} vortices with a fractional magnetic flux replaces the so-called LO vortex lattice \cite{I2,HM} with periodic nodal planes perpendicular to the field ${\bf H}$ and have argued that thermal conductivity $\kappa$'s data \cite{Capan} are consistent with the former state rather than the latter. Below, it is pointed out that available experimental facts in CeCoIn$_5$ are incompatible with properties peculiar to the former state  and that the analysis in Ref.\cite{Agter} is lacking in a firm theoretical basis. 

First, based on their notation \cite{Agter}, the state they propose is more stable than the ordinary LO vortex lattice \cite{I2,HM} if $2 {\tilde \beta}_1 \beta_A(0)/\beta_A({\bf \tau}) > 2 {\tilde \beta}_1 + {\tilde \beta}_2 > 0$. When $\eta_1=\eta_2$, this condition implies assuming the case with a second order $H_{c2}(T)$-transition. Opposite to this, the $H_{c2}$-transition in CeCoIn$_5$ at lower temperatures is discontinuous \cite{Bianchi,Watanabe}, implying that, according to their criterion, the LO vortex state is stable as the high field phase. In fact, the structure of their phase \cite{Agter} is described by a couple of order parameters, $q$ and ${\bf \tau}$, defining differences in structures from the Abrikosov lattice with $q={\bf \tau}=0$, and thus, the structural change between the crisscrossing and Abrikosov lattices needs to occur through a couple of transitions or a single discontinuous one in contrast to a single continuous transition in CeCoIn$_5$ \cite{Bianchi,Watanabe,Capan,Miclea}. Besides this, they have assumed the validity of the perturbative expansion in the gradient {\it perpendicular} to ${\bf H}$ in their starting model. This gradient expansion is valid only in the case with a large enough Maki parameter \cite{HM} where the $H_{c2}(T)$ and the vortex state at lower temperatures are described in the ballistic (or clean) limit by the SC order parameter in not the lowest Landau level (LL), assumed in Ref.\cite{Agter}, but a higher LL \cite{AI}. In the next lowest LL state, periodic nodal lines appear with a distance of the order $(\sqrt{2 e H})^{-1}$ in the plane perpendicular to ${\bf H}$, and hence, the transition between such a higher LL state and the ordinary Abrikosov lattice is inevitably of first order in contrast to the observation in CeCoIn$_5$. We note that, according to Fig.7 of Ref.\cite{AI}, a modulation parallel to ${\bf H}$ does {\it not} occur in the higher LL. 

One of key experiments reflecting the structure of the high field state is the ultrasound measurement \cite{Watanabe}, in which the sound velocity has remarkably decreased upon entering the high field phase from the Abrikosov state only in the Lorentz mode with a displacement perpendicular to ${\bf H}$. This has been explained \cite{I2} as an evidence of the presence of nodal planes perpendicular to ${\bf H}$. On the other hand, the crisscrossing lattice represented in the lowest LL \cite{Agter} has no nodal planes, and the vortices, which are not parallel to ${\bf H}$, will be strongly pinned by the {\it crystal} lattice even in the non-Lorentz mode. Thus, this state realized as the high field state would result in an enhancement of sound velocity in contrast to the observation \cite{Watanabe}. 
Further, the fact \cite{Miclea} that the discontinuous $H_{c2}$-transition and the high field phase appear over {\it wider} temperature ranges at a {\it higher} pressure implies that a magnetic ordering is inessential to them. Finally, the observed feature \cite{Capan} noted in Ref.\cite{Agter} that $\kappa$ parallel to ${\bf H}$ is greater than that perpendicular to ${\bf H}$ is also seen clearly in the Abrikosov state in lower fields, and thus, no one can regard it as a feature peculiar to the high field phase. The increase of $\kappa$ parallel to ${\bf H}$ in the high field phase originates from a decrease of the amplitude of SC order parameter \cite{I2} and does not contradict the picture that the high field phase is the LO 
vortex lattice with nodal planes perpendicular to ${\bf H}$ \cite{neutron}. 

In conclusion, the high field phase of CeCoIn$_5$ is not the crisscrossing lattice \cite{Agter}, although the proposed state might appear in future in other materials.

\vspace{0.3truecm}
\bigskip

\noindent Ryusuke Ikeda \\
Department of Physics, Graduate School of Science, \\
Kyoto University, Kyoto 606-8502, Japan 

\vspace{0.3mm}

\end{document}